# A Hybrid Rule-Based and AI-Augmented Framework for Automatic Failure Recovery in DevOps Deployments

Raju Chowdhury
*Independent Researcher*
Seattle, WA, USA
raju.chowdhury.1384@gmail.com

Aniruddha Singh Gautam
*Independent Researcher*
Seattle, WA, USA
aniruddhasinghg81@gmail.com

Apurv Ghai
*Independent Researcher*
Dallas, TX, USA
apurvgh@gmail.com

***Abstract*—Automatic failure recovery is another difficult area of DevOps deployments due to the complexity of the system, high workload and the limitations of the traditional rules-based or manual approach. This study proposes a hybrid model integrating deterministic, rule-based recovery with the assistance of machine learning by fault prediction to automatically monitor failures, identify, classify and recover failures in real time. To test the architecture, a distributed log dataset of 100,000 records was used for models such as Decision Tree, Random Forest, Logistic Regression, LightGBM, Autoencoder with BiLSTM. The Decision Tree had a strong performance in defect identification with an F1 score of 80.8%, recall of 80.0%, precision of 81.6%, and accuracy of 89.4%. An impressive 83.3% success rate was achieved by the automated recovery measures, resulting in a 94.6% decrease in mean time to recovery (MTTR), a 95.4% reduction in downtime, and annual savings of 7,462. The combination of explainable rule-based logic and adaptive AI prediction in the framework means it is more resilient, operationally efficient, and scalable, and an effective, autonomous approach to DevOps environments evaluated under controlled experimental conditions.**

## I. INTRODUCTION

The evolution of digital technologies is increasing very fast and has changed the dynamics of software development, deployment, and maintenance in industries [1], [2]. Traditional patterns of software delivery were built on sequential development and manual deployment, which were slow, error-prone, and unable to keep up with current high-frequency release cycles. Agile solutions served to address part of these issues through the iterative development process and continuous integration maintenance [3], [4]. To fill the gap between operations and development, the DevOps concept was created to conquer the absence of connection between the software development stage and the deployment of the same software into production within large software organizations [5][6]. To address these vulnerabilities, automatic systems that recover failures have been proposed that utilize real-time checks, warnings, and healing systems to reduce downtime and enhance the resilience of the system. Fault prediction using machine learning and reinforcement learning are two cases of AI-powered approaches that have demonstrated potential in detecting the complicated nature of the failure and supporting adaptive recovery [7], [8]. To overcome the shortcomings, this paper presents a hybrid rule-based and solution to automatic failure recovery, a combination of deterministic recovery logic in case of predictable failures and fault prediction in case of unknown or dynamic failures [9]. This paper is motivated by the shortcomings of current AI- and rule-based recovery systems in DevOps. AI augmentation in this work refers to the integration of machine-learning-assisted fault prediction within an otherwise deterministic rule-based recovery pipeline. Solutions that rely solely on AI are not necessarily explainable, require a significant amount of labeled data and are rule-based, generally needing human involvement. The framework encompasses both rule-based and AI-driven fault prediction methods and is designed to boost operational efficiency, minimize downtime, and fortify system resilience in large-scale implementations. The key research findings are presented below:

- Defining automated failure recovery using a combination of rules and AI for DevOps deployments.
- Deterministic recovery logic and machine learning-based fault prediction for autonomous and adaptive recovery.
- Significant analysis of predictive, operational and economic indicators that indicate the performance of MTTR, downtime, recovery success and system stability has improved.
- Scalability, explanation and practical solution provision to DevOps pipelines at scale.
- Evidence of cost-effectiveness with recognizable operational cost savings, and short payback of ROI.

### A. Structure of Paper

The paper is structured in following way: **Section II** is an overview of related work. A suggested hybrid framework and approach are explained in **Section III**. The experimental setup and results are given in **Section IV**. Lastly, **Section V** concludes research and describes future research directions.

## II. LITERATURE REVIEW

The literature emphasizes the progress in AI for DevOps recovery, including LLMs and machine learning, offering limitations of the existing approaches. R. Dhawan and M. Dhawan (2026) This paper presents a smart CI/CD framework using a Sense–Analyze–Predict–Act–Learn (SAPAL) loop for adaptively and reliably executing the pipeline. The metrics make it easier to deliver, and false failure prevention reduces the number of false failures, while ML-based test selection optimizes delivery efficiency [10]. Yang et al. (2025) Introduce Intelligent Fault Self-Healing Mechanism (IFSHM) based on Large Language Models (LLM) and Deep Reinforcement Learning (DRL). The experiment proves that IFSHM allows achieving a 37% reduction in the recovery time of the system [11].

Brahmandam, Punjabi and Chandramohan (2025) present AI-enhanced DevOps by integrating agents based on LLM

and machine learning. Simulated CI/CD experimental results indicate that deployment frequency increases by 1 to 4.8/day, MTTR decreases by 183 to 38 minutes, and the change failure rate decreases by 21 to 5.6% [12]. Madabushini (2025) work evaluates an AI-based and multi-agent self-healing infrastructure applied in DevOps environments. AI-based automation allows 24-hour surveillance, shortening response time to 45 minutes, and improving recovery performance with a response time of 15 minutes, a 95% detection rate, and 90% automated recovery [13].

Desmond (2024) explores AI-enhanced DevOps systems that combine predictive analytics and self-healing capabilities. The experimental outcomes show a significant improvement in traditional DevOps such as MTTR by 41.7%, a higher release velocity by 40% and higher accuracy of anomaly detection by 70-91% [14]. Kim, Choi and Jung (2024) propose an automated recovery system that enables the use of Kubernetes management and backup tools. Testing of ten failure modes showed that the average recovery time of 27 seconds was obtained [15].

**Research Gap:** Despite advances in AI-augmented DevOps and self-healing infrastructures, current approaches exhibit key limitations. The majority of frameworks are based on AI or LLM-based recovery and usually need human intervention and lack transparency or governance, where deterministic rule-based protections are not fully applied. Also, the assessment is often performed within simulated or constrained cloud setups and can only be used in scaled-up DevOps pipelines. The majority of methods involve predictive fault detection, as opposed to end-to-end autonomous recovery, and general recovery rules are not applicable to new or unexpected failures. These limitations validate the need for a hybrid solution, where rule-based and AI-based automation are combined to achieve reliable, transparent and completely autonomous failure recovery for real DevOps deployments.

## III. Methodology

The paper proposes an integrated rule-based and AI-enhanced system for automatic failure recovery in DevOps, as shown in Fig. 1. The framework consists of two main stages: Model Development & Deployment and Real-Time Monitoring & Recovery. In the first stage, Log Data Collection gathers system logs, followed by Data Preprocessing, which includes missing value checking, timestamp normalization, duplicate removal, feature extraction, feature engineering, and label encoding. The dataset is then split into Training (70%) and Testing (30%). Further Preprocessing is applied using SMOTE for data balancing and StandardScaler for feature scaling. Multiple Model Learning Models, such as LR, DT, RF, LightGBM, and Autoencoder with BiLSTM, are trained and evaluated using Evaluation Metrics including accuracy, precision, recall, and F1-score. The best model is selected through Best Model Selection & Serialization and deployed via Deployment. In the second stage, Real-time Log Monitoring pushes logs to the AI-Fault Prediction Engine which then pushes to the Rule-Based Classification Engine, the Failure Mapping and Recovery Action Selection, and Automated Recovery Execution modules, which are monitored through the Performance Monitoring, Impact Analysis and Reporting modules for continuous operational assessment.

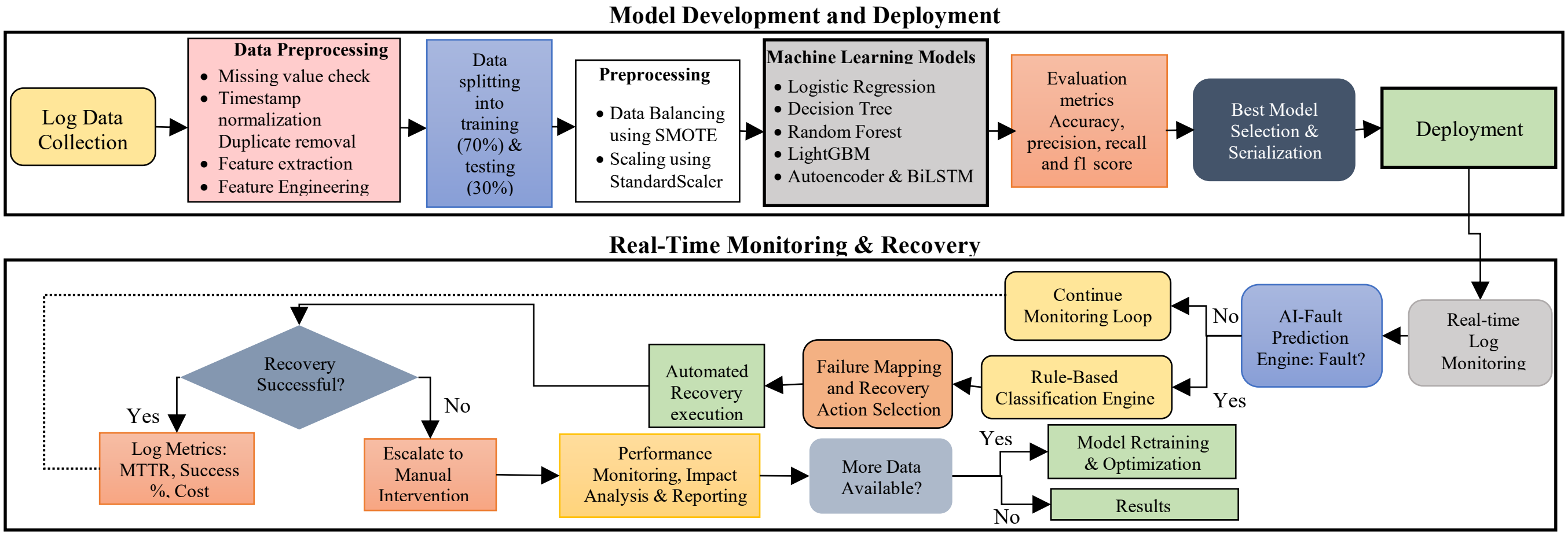

Fig. 1. Workflow of Hybrid AI and Rule-Based Failure Recovery Framework.

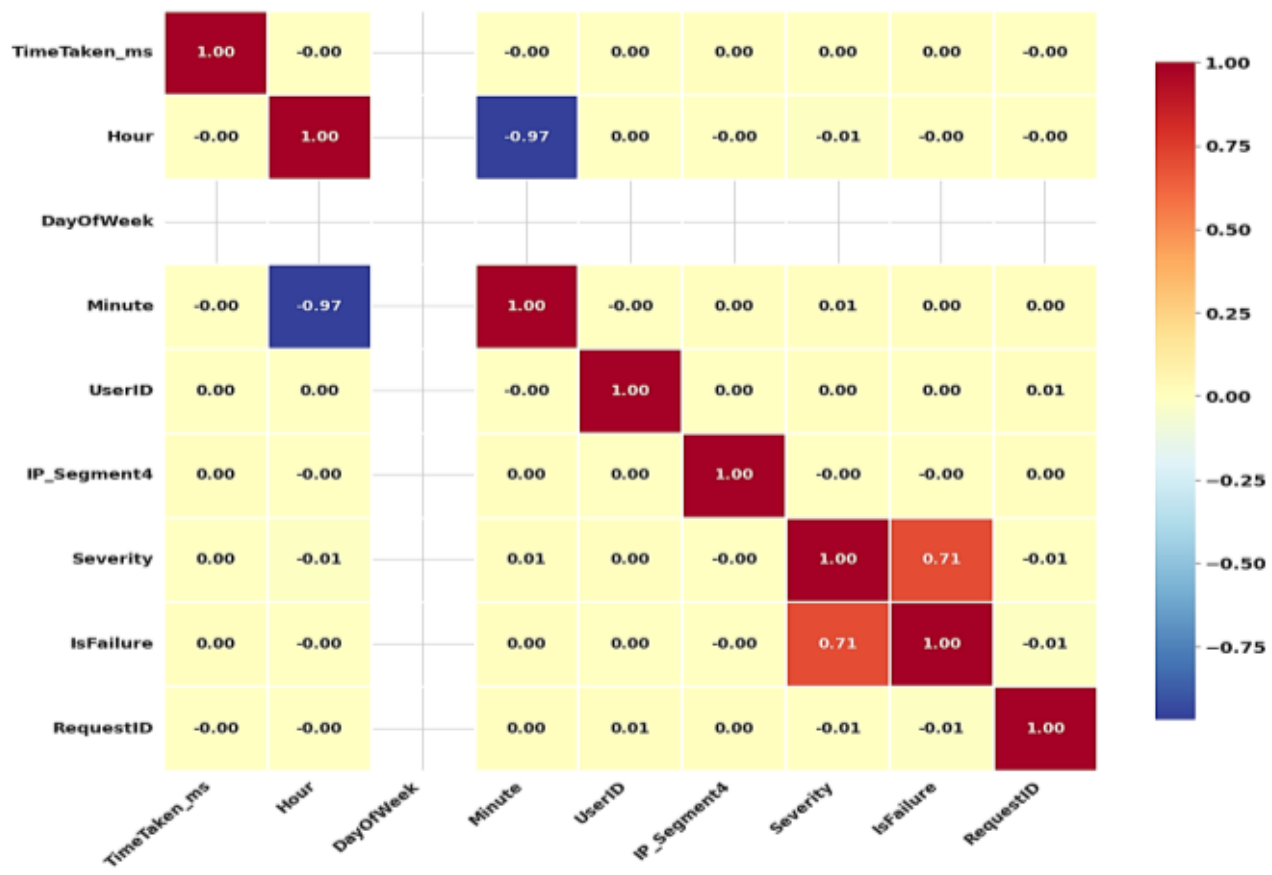

Fig. 2. Correlation heatmap of dataset attributes and engineered features.

## A. Data Gathering

The proposed framework uses the Synthetic Log Data of Distributed System dataset [16] which is loaded from logdata.csv. It has 100,000 log records with nine attributes, including timestamp, log level, service name, message type, request ID, user identifier, client IP address, and request execution time. During preprocessing, additional engineered features such as Severity, IsFailure, CPU utilization, memory usage, latency, failure frequency, and recovery indicators were derived from the original log fields to improve fault prediction. Fig. 2 presents the correlation heatmap of both the original and engineered features used for model training.

## B. Data Preprocessing

The raw log data is pre-processed in order to ensure reliability, consistency, and appropriateness for predictive modelling. The steps are provide:

### 1) Missing-Value Handling

Checks whether a record or a part of a record is missing or not complete and deletes or imputes it to ensure reliability.

### 2) Timestamp Normalization

Converts and standardizes the datetime fields into a consistent datetime format that allows for a proper temporal sequence and analysis of failure and recovery events.

### 3) Duplicate Removal

Removes duplicate and inconsistent data from re-monitoring or redeploying the same code into the system, reducing noise in the dataset.

### 4) Feature Engineering

There are nine raw attributes in the original dataset. To improve predictive performance, additional engineered features, such as Severity, IsFailure, failure frequency, downtime, recovery time, success indicators and system state flags, were created during preprocessing.

### 5) Feature Extraction

Converts raw log fields (events, error codes, component IDs, timestamps, recovery actions, outcomes) into structured variables for modelling.

### 6) Label Encoding

Label encoding is the approach used to convert categorical attributes to numerical values. The different categories (e.g. log level, service name) obtain a distinct integer based on a LabelEncoder in this method and machine learning models can process categorical features of logs successfully.

## C. Data Splitting

A 70% training set and a 30% testing set comprise data. The ML models are trained with the training set, and tested on the testing set.

## D. Data Balancing Using SMOTE

In the training phase, the Synthetic Minority Oversampling Technique (SMOTE) is used to balance class frequencies (failure and non-failure). This improves model's ability to properly diagnose rare yet severe failure states.

## E. Scaling Using StandardScaler

The preprocessing technique StandardScaler guarantees feature normalization by removing the mean and adjusting to unit variance. This increases model convergence by giving each feature a 0 mean and a unit standard deviation. The transformation is defined by Equation (1):

$$z = \frac{x-\mu}{\sigma} \quad (1)$$

Where $z$ is standardized value, $x$ is initial feature value, $\mu$ is the feature mean, and $\sigma$ as the standard deviation.

## F. Machine Learning Models

A number of ML models, such as LR, DT, RF, LightGBM, and an Autoencoder with Bidirectional Long Short-Term Memory (Bi-LSTM) network, are trained and assessed for defect prediction. Among these, DT model is described, and its effectiveness in fault classification is discussed.

A DT is a hierarchical, nonparametric model of learning that splits feature space into nonparametrically homogeneous regions recursively. The tree has internal decision nodes and terminal leaf nodes with each internal node $n$ applies a decision function $f_n(d)$ to an input sample d to determine the traversal path. For a node n, entropy is defined in Equation (2):

$$H(n) = -\sum_{i=1}^{N} p_n^i log_2 p_n^i \quad (2)$$

Where H(n) represents the entropy of node n, N is the number of classes (here N=2, representing the normal and failure classes), and $p_n^i$ denotes the probability of class Ci (0 or 1) at node n, given that a data sample reaches that node. The information gain function G for feature $x_i$ at node n is then defined by Equation (3).

$$G(x_i, n) = H(n) - H(x_i, n) \quad (3)$$

Where $H(x_i, n)$ denotes the sum of entropy of children nodes after splitting node n based on feature $x_i$.

### 1) Model Training

The DT model is trained by reducing node impurity using the Gini Index as the primary splitting criterion. The maximum depth of tree is set to 10 to avoid overfitting model, and the minimum number of samples to split and 2 to occupy a leaf node is 5, respectively. Cost-complexity pruning with a regularisation parameter α=0.01 is used for post-pruning. Within these constraints, the decision tree learns the best decision rules by recursively partitioning the feature space to distinguish between normal and failure states.

## G. Best Model Selection

The Decision Tree was selected because it achieved the highest classification performance while maintaining high interpretability. Its transparent decision rules enable direct mapping of predicted faults to recovery actions, making it more suitable for automated DevOps environments than more complex ensemble models.

### *H. Failure Detection and Recovery Process*

The selected model is deployed in production, which is the trained machine learning model, as an AI-based fault prediction engine. The process steps are as follows:

- **Real-Time Log Monitoring System:** This system records production logs every minute, which are used to identify failures in real time, such as events, resource usage, service conditions and errors.
- **Machine Learning Fault Prediction:** The deployed ML model analyzes incoming system logs and predicts potential failure events in real time.
- **Monitoring Loop:** Stable behavior and recovery results are monitored and anomalies or frequent failures are detected in order to make adaptive responses.
- **Rule-Based Failure Classification:** Predicts the faults based on the pre-defined rules and each fault is correlated with the corresponding recovery action.
- **Recovery Action Selection:** Selects and automatically initiates the most suitable recovery plan for returning to normal operation, while minimizing downtime.
- **Automated Recovery Execution:** The selected recovery action is automatically executed, thereby reducing the downtime of the system and enhancing service continuity. Recovery rules are applied only after fault prediction for recovery mapping, not during training.
- **Performance Monitoring, Impact Analysis & Reporting:** Monitors and assesses performance measures, such as MTTR, downtime, and recovery success to make decisions.
- **Model Retraining & Optimization:** The model is retrained with new data periodically and optimized to keep its accuracy in terms of prediction.

### *I. Evaluation Metrics*

The proposed system is evaluated with the classification, recovery, operational and economic metrics. To evaluate the recovery success rate and recovery time in the process of automated fault handling, the recovery success rate and recovery time are employed. To evaluate the classification performance, accuracy, precision, recall and F1-score are employed [17]. The system's stability, average downtime, mean time to recover (MTTR), and failure count are all indicators of its reliability. Economic impact is measured by cost savings and ROI payback period. These metrics are defined mathematically in Equations (4)-(9).

$$Accuracy = \frac{TP+TN}{TP+TN+FP+FN} \quad (4)$$

$$Precision = \frac{TP}{TP+FP} \quad (5)$$

$$Recall = \frac{TP}{TP+FN} \quad (6)$$

$$F1 = 2 \times \frac{Precision \times Recall}{Precision+Recall} \quad (7)$$

$$Avg\ Downtime = \frac{1}{N_f}\sum_{i=1}^{N_f} D_i \quad (8)$$

$$MTTR = \frac{1}{N_f}\sum_{i=1}^{N_f}(t_{r,i} - t_{f,i}) \quad (9)$$

Here, TP, TN, FP, and FN denote true positives, true negatives, false positives, and false negatives, respectively, while $N_f$ represents the total number of failure events. The variables $t_{f,i}$ and $t_{r,i}$ indicate the failure occurrence time and recovery completion time of the i-th failure, respectively.

## IV. Experimental Results and Analysis

This paper assesses a hybrid rule-based and AI-enhanced framework for the automatic recovery of failures in DevOps deployments. The proposed framework was evaluated using the publicly available Log Data of Distributed System dataset from Kaggle. Since the dataset is synthetic, the experimental results should be considered a proof-of-concept validation under controlled conditions. The experiments were conducted using Python 3.11 on Ubuntu 22.04 LTS with an 8-core 3.0 GHz CPU, 32 GB RAM, and SSD storage. The preprocessing pipeline, feature engineering, model configurations, hyperparameters, and evaluation protocol are described to facilitate reproducibility, while the implementation is available through the accompanying GitHub repository[18]. Table I summarizes the characteristics of the simulated faults, including CPU usage, memory consumption, service latency, error frequency, and the corresponding recovery actions.

TABLE I. Characteristics of Simulated Faults in DevOps Pipelines

| Fault Type | CPU Usage | Memory Usage | Latency | Error Frequency | Recovery Action |
|---|---|---|---|---|---|
| CPU_OVERLOAD | >80% | 60–70% | >800 ms | >5 errors/min | Restart Service |
| MEMORY_LEAK | 50–70% | >85% | 400–800 ms | 2–5 errors/min | Clear Cache |
| SERVICE_CRASH | 30–70% | 40–80% | 10–100 ms | FATAL logs | Redeploy Application |
| LATENCY_SPIKE | <60% | <60% | >1000 ms | 1–3 errors/min | Auto-Scale Resources |

TABLE II. Performance of the AI model for Failure Detection

| Model | Accuracy (%) | Precision (%) | Recall (%) | F1-Score (%) |
|---|---|---|---|---|
| Logistic Regression | 78.5 | 42.3 | 74.8 | 54.1 |
| Decision Tree | 89.4 | 81.6 | 80.0 | 80.8 |
| Random Forest | 87.2 | 76.8 | 78.2 | 77.5 |
| LightGBM | 86.9 | 76.1 | 77.8 | 76.9 |
| Autoencoder | 68.3 | 24.8 | 43.7 | 31.6 |
| BiLSTM | 86.5 | 75.4 | 77.0 | 76.2 |

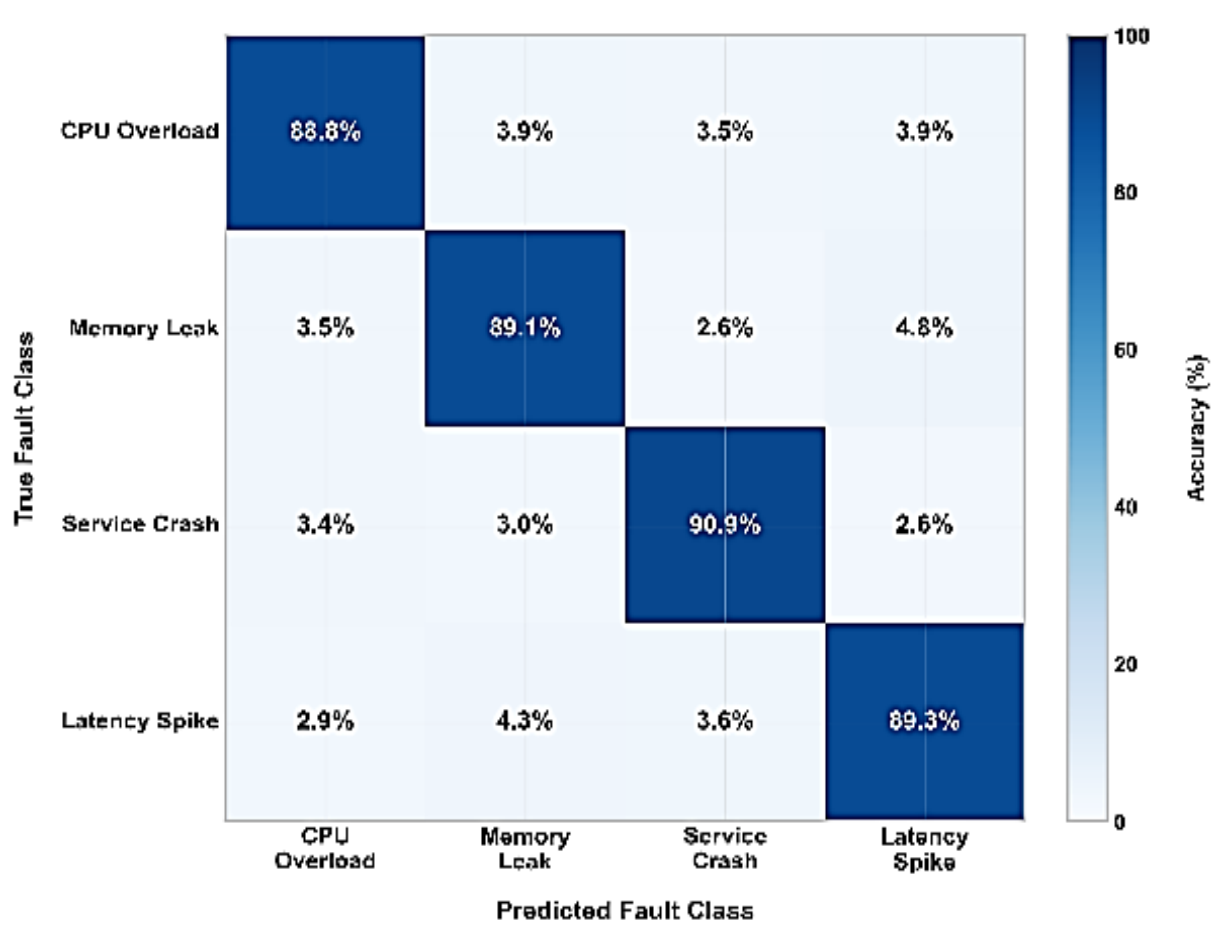

Fig. 3. Normalized Confusion Matrix for Decision Tree Fault Classification.

As shown in Table II, which compares all trained machine learning models using the same train-test partition and preprocessing pipeline, the DT model achieved the high classification performance, with 89.4% accuracy and balanced precision, recall, and F1-score. The performance of RF, LightGBM and BiLSTM was competitive, but slightly lower than that of the NN. The normalized confusion matrix of the Decision Tree, on test data, is shown in Fig. 3. As can be seen from this matrix, the model is able to correctly classify CPU overload, memory leak, service crash and latency spike. A fault-injection simulation reports the success rate of automated redeployment to be 90% and that of cache clearing to be 75%, for a total recovery success of 83.3% (Table III) when comparing the number of predicted fault-driven redeployments with the expected number of redeployments in the fault–response mapping table.

TABLE III. DEVOPS AUTOMATED RECOVERY STRATEGY PERFORMANCE

| Failure Type | Recovery Action | Success Rate (%) | Execution Time (s) |
|---|---|---|---|
| CPU Overload | Restart Service | 85 | 30 |
| Memory Leak | Clear Cache | 75 | 15 |
| Service Crash | Redeploy App | 90 | 120 |
| Latency Spike | Auto-Scale | 80 | 60 |
| Overall 83.3% | | | |

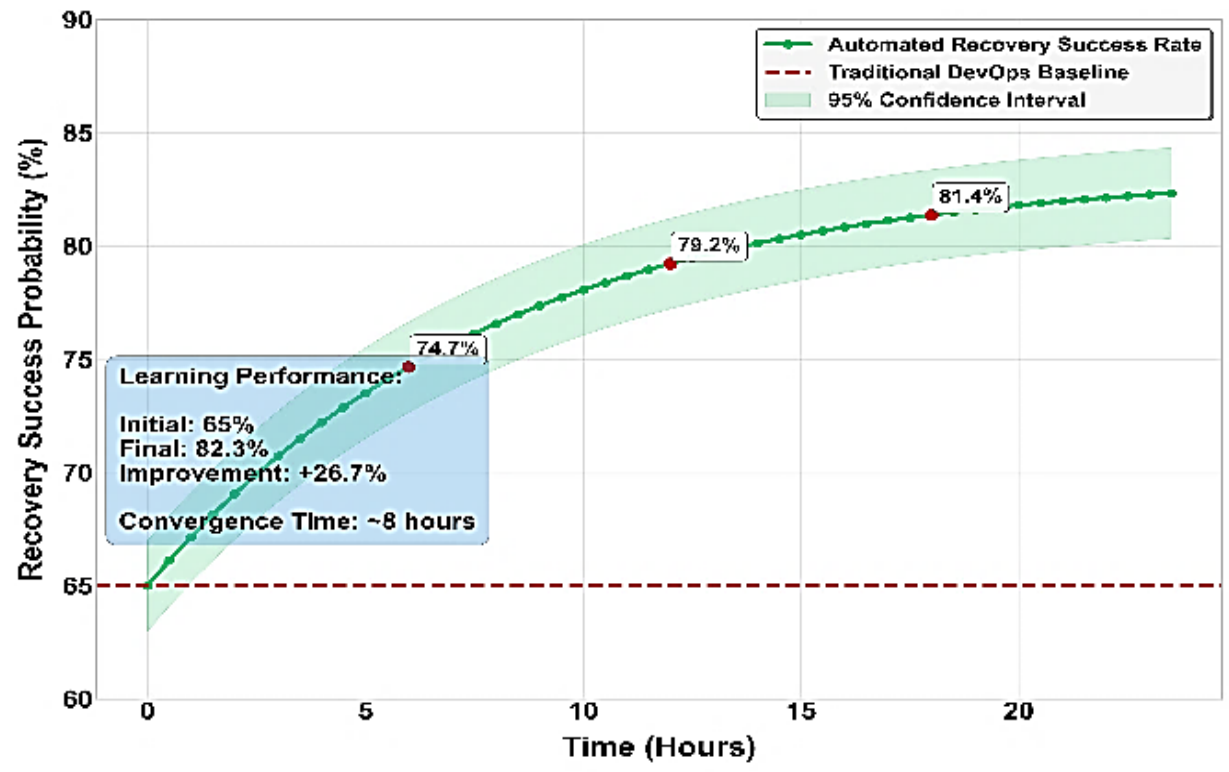

Fig. 4. Performance Improvement Across Retraining Cycles.

Fig. 4 illustrates the model performance after periodic batch retraining using newly accumulated log records. The framework does not employ online or continual learning. Instead, the model is retrained offline at scheduled intervals, and the observed performance improvements reflect incremental incorporation. Table IV compares hybrid system with the traditional system, a static, rule-based monitoring system widely used in DevOps environments. It does not use ML and only works with pre-programmed rules for fault detection and recovery, resulting in over 94.6% reduction in failures, downtime and MTTR, 28.2% improvement in recovery success, 4.3% improvement in system stability and a total of 7,462 dollars saved per year, which is both efficient and cost-effective in relation to operations.

TABLE IV. COMPARISON OF PROPOSED SYSTEM WITH RULE-BASED BASELINE SYSTEM

| Metric | Rule-based Baseline System | Hybrid System | Improvement |
|---|---|---|---|
| Failure Count | 27,776 | 1,000 | 96.4% ↓ |
| Average Downtime | 25.5 min | 1.2 min | 95.4% ↓ |
| MTTR | 18.3 min | 1.0 min | 94.6% ↓ |
| Recovery Success | 65.0% | 83.3% | 28.2% ↑ |
| System Stability | 84.7% | 88.4% | 4.3% ↑ |
| Annual Cost Savings | — | $7,462 | 3–6 month ROI |

The baseline used for comparison is a deterministic rule-based monitoring system without machine learning. The baseline continuously monitors predefined thresholds for CPU utilization, memory usage, latency, and error frequency, using the same fault definitions. Recovery actions are triggered whenever predefined thresholds are exceeded. Both the rule-based baseline and the proposed framework were evaluated using the identical synthetic dataset, preprocessing pipeline, recovery rules, and evaluation protocol. Therefore, the comparison isolates the contribution of machine-learning-assisted fault prediction while keeping the recovery policy unchanged.

Economic impact was estimated using the simplified cost model described above. The reported annual savings ($7,462) and 3–6 month ROI are based on assumed downtime and labor costs for comparative evaluation under the experimental setup, rather than organization-specific financial data.

Fig. 5 shows the MTTR distribution for the conventional DevOps model and the hybrid model with AI. The hybrid method shows smaller and predictable time to recovery, and the method is efficient in dealing with failures and decreasing operating times.

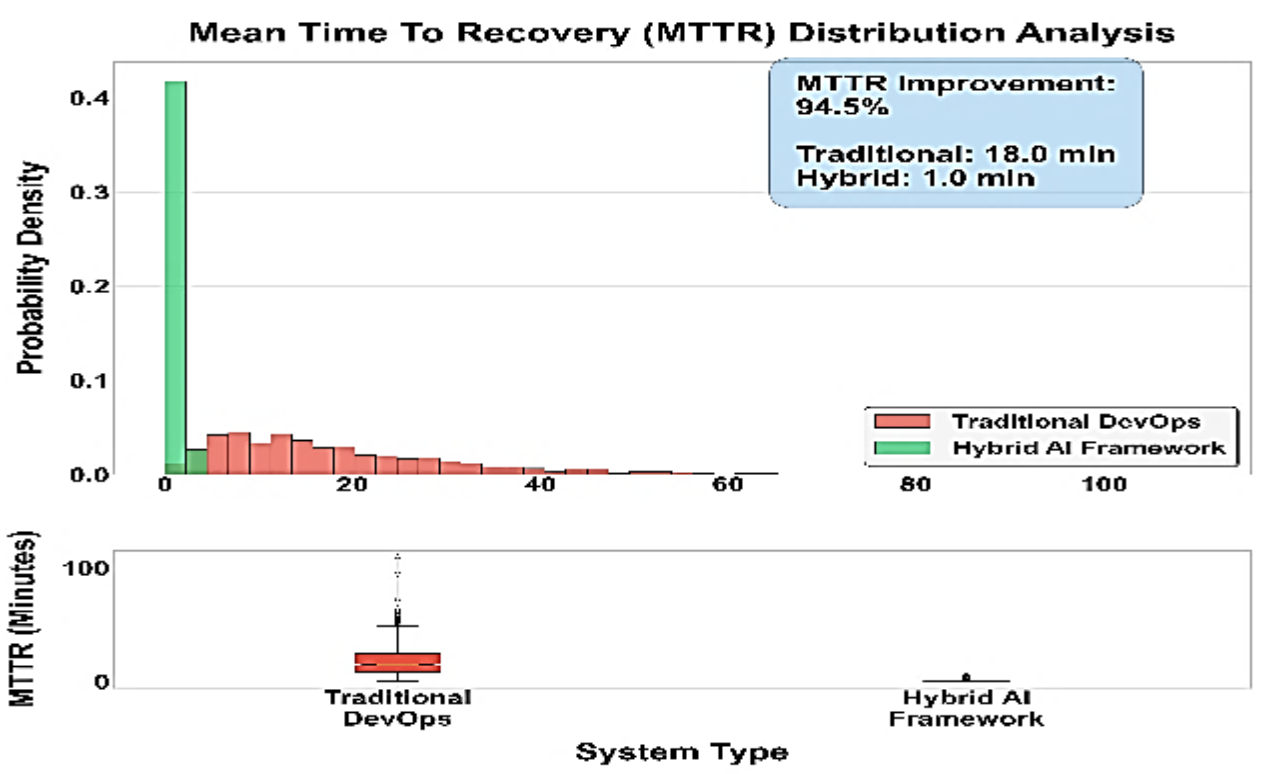

Fig. 5. Mean Time to Recovery (MTTR) Distribution Analysis.

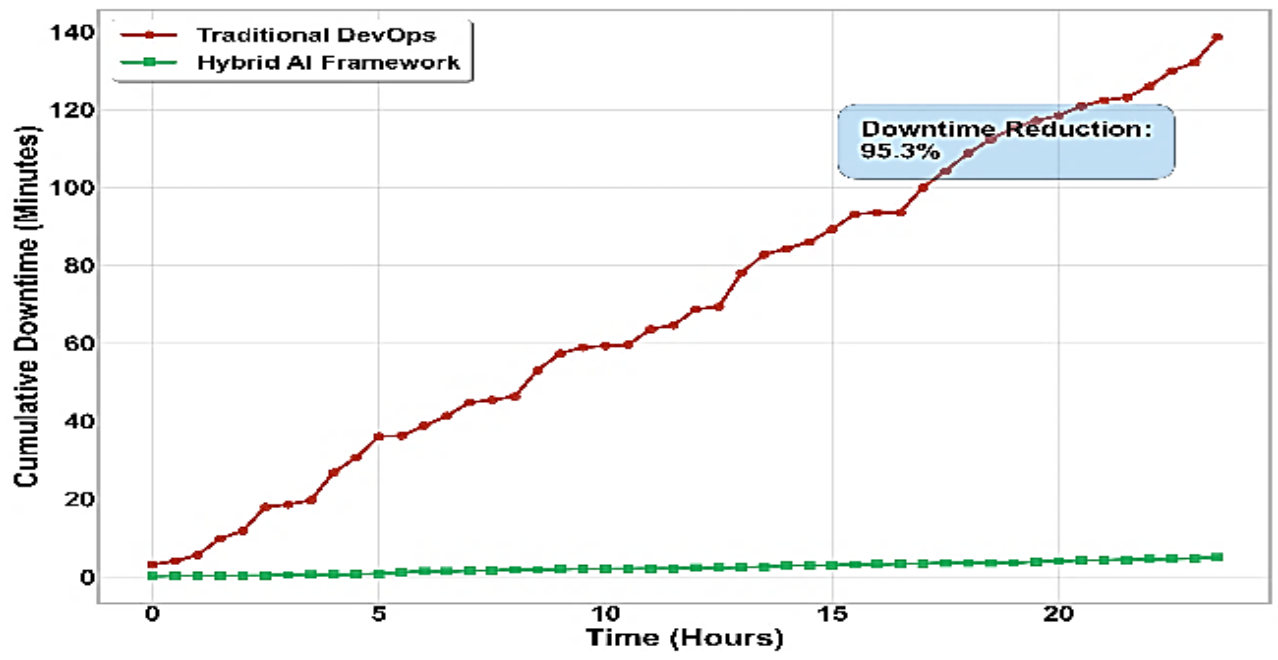

Fig. 6. System downtime comparison of AI and traditional frameworks.

The comparison in Fig. 6 shows the 24 Hour cumulative downtime for both the conventional and hybrid AI-augmented framework, and the hybrid framework seems to significantly minimize the downtime, improving the recovery rate and system resilience. Fig. 7 gives the availability of the systems (in comparison to the SLA standards) and the hybrid solution always maintained a level at which it exceeded the standards, making this approach suitable for mission-critical DevOps.

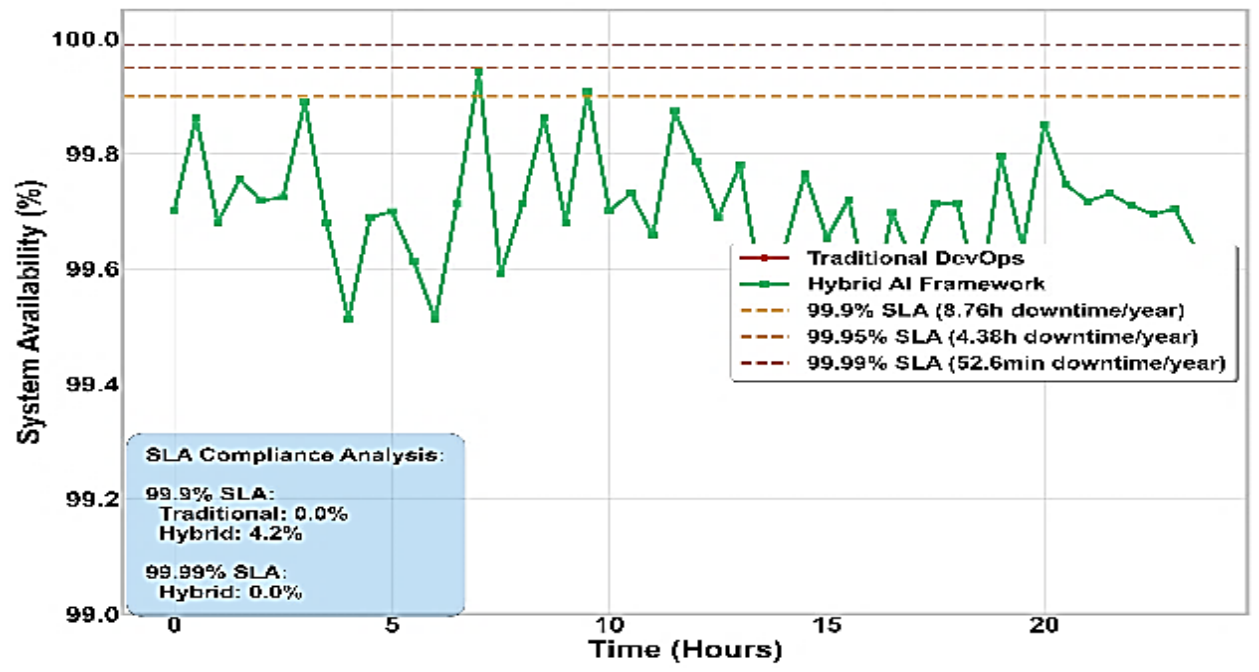

Fig. 7. System Availability Comparison with Industry SLA Standards.

## *A. Discussion*

The discussion section explains interpretation of the experimental outcomes and why proposed hybrid framework is able to detect failures, recover faster than traditional DevOps automation, and be more resilient. Although RF and LightGBM typically outperform a single DT, they showed no clear advantage on this dataset. The use of the Decision Tree model resulted in good predictive accuracy, both for classifying the fault with high confidence and for implementing effective recovery measures. It should be noted that the synthetic benchmark contains predefined fault characteristics that simplify the classification task compared with real production environments. Therefore, reported predictive performance demonstrates feasibility of integrating machine learning with rule-based recovery rather than establishing superiority under operational DevOps conditions.

**Limitations:** A few limitations of the present study should be noted. A synthetic distributed system log dataset was used to evaluate the proposed framework in controlled environments, but it may not accurately capture the complexity, heterogeneity and dynamic nature of actual DevOps systems. Moreover, only four categories of faults were evaluated (CPU overload, memory leak, service crash, and latency spike), which limits the generalizability of the results to other failure scenarios such as network partitions, cascading dependency failures, configuration drift, security failures, and heterogeneous workload failures. Furthermore, there was a lack of proper statistical validation of the experimental evaluation, such as repeated trials, k-fold cross-validation, confidence intervals, and significance testing. Thus, the results can be viewed as proof-of-concept validation and not as actual production deployment results.

## V. Conclusion and Future Study

The proposed hybrid rule-based and AI-enhanced framework has successfully overcome the limitations of the traditional rule-based and AI-only systems of automatic failure recovery in the DevOps environment. The combination of machine learning-based fault prediction and deterministic recovery mechanisms allows for efficient fault detection, classification, and remediation with minimal human intervention. The experimental results show increased prediction accuracy, higher recovery rates, more efficient operations, and lower MTTR and downtime when using a hybrid intelligent automation approach compared with conventional approaches for DevOps systems, thereby supporting the feasibility of hybrid intelligent automation in DevOps systems. Furthermore, framework brings explainability and reliability to recovery actions and decision-making processes. While these results are promising, there is a lack of fault diversity and statistical power. Other, more complex, and realistic fault categories, such as network failures, cascading dependencies, configuration drift, and security-related faults, will be added to the framework to make it more scalable and applicable to real-world scenarios. Furthermore, future studies will involve thorough statistical testing with multiple experimental runs, k-fold cross-validation, confidence intervals and significance testing to improve reliability and reproducibility. Future developments could include reinforcement learning techniques for adaptive recovery optimization, coordination between multiple agents, and an interpretive capability for semantic logs using an LLM for enhanced fault diagnosis and remediation.